\documentstyle[12pt,aaspp4]{article}
\newcommand{\logg}{$\log g$}

\newcommand{\lya}{\mbox{Ly$\alpha$}}
\newcommand{\mgii}{\ion{Mg}{2} }

\newcommand{\msun}{M$_{\sun}$}

\newcommand{\rsun}{R$_{\sun}$}
\newcommand{\teff}{T$_{\rm eff}$}

\slugcomment{06 Mar 1997, accepted by ApJL}
\begin{document}  

\title{S1040 in M67: A Post-Mass-Transfer Binary with a Helium-Core 
White Dwarf} 

\author{W. Landsman\altaffilmark{1},
J. Aparicio\altaffilmark{2}, 
P. Bergeron\altaffilmark{3}, 
R. Di Stefano\altaffilmark{4},
T. P. Stecher\altaffilmark{5} }

\altaffiltext{1}{Hughes STX Corporation, NASA Goddard Space Flight Center,
Laboratory for Astronomy and Solar Physics, Code 681, Greenbelt, MD 20771}

\altaffiltext{2}{D\'{e}partement de Physique, Universit\'{e} de Montr\'{e}al,
Succ.\ Centre-Ville, C.P. 6128, Montr\'{e}al, Qu\'{e}bec, Canada, H3C~3J7}

\altaffiltext{3}{Lockheed Martin Canada, 6111 av. Royalmount, 
Montr\'{e}al, Qu\'{e}bec, Canada, H4P~1K6}

\altaffiltext{4}{Harvard-Smithsonian Center for Astrophysics, 60 Garden Street,
Cambridge, MA 02138}

\altaffiltext{5}{NASA Goddard Space Flight Center,
Laboratory for Astronomy and Solar Physics, Code 681, Greenbelt, MD 20771}

\vspace{0.1in}

\begin{abstract}

We have obtained spectra of the yellow giant S1040 in the
open cluster M67 using the Goddard High-Resolution Spectrograph (GHRS)
and the Faint Object Spectrograph on the {\em Hubble Space Telescope}.
S1040 is a single-lined spectroscopic binary with a 42.8d period that 
occupies a ``red straggler'' position in the M67 color-magnitude diagram (CMD),
0.2 mag blueward of the giant branch.
A detection of S1040 at 1620 \AA\ with the {\em Ultraviolet Imaging 
Telescope} provided evidence that the secondary is a hot white dwarf, 
and thus that the anomalous location of S1040 in the CMD is likely 
due to a prior episode of mass-transfer.   Our GHRS spectrum shows
a broad \lya\ absorption profile that confirms the white dwarf identification 
of the S1040 secondary.
A model atmosphere
fit to the GHRS spectrum yields \teff\ = 16,160 K, \logg\ = 6.7, and a mass of 
$\sim 0.22$ \msun, for an assumed cluster 
distance of 820 pc and reddening of E(B--V) = 0.02.     The unusually low mass 
derived for the white dwarf implies that it must have a helium core, and 
that a mass-transfer episode
must have begun while the progenitor was on the lower
giant branch.    We construct a plausible 
mass-transfer history for S1040 in which it originated as a short ($\sim$ 2d) 
period binary, and evolved through a blue straggler phase to reach its 
current state.

\end{abstract}


\section{Introduction}                                                       

 S1040 (catalogued by \cite{sand77}; =Fagerholm 143) is a yellow giant with 
 \bv = 0.86, and an astrometric and radial
 velocity  member of the open cluster M67 (\cite{gir89}; \cite{mlg90}).  
 In their radial velocity study of M67, \markcite{mlg90}Mathieu et al.\ 
 discovered that S1040 was a single-lined  spectroscopic binary with a 
 circular orbit and a  period of 42.8 days.
 In the M67 color-magnitude diagram (CMD),  S1040 occupies a ``red straggler'' 
 position, 0.2 mag blueward of the giant
 branch.    \markcite{js84}Janes and Smith (1984) suggested that this 
 anomalous location of S1040 in the CMD could be explained if S1040 were
 a photometric binary consisting of a star on the lower giant branch and
 a star near the main-sequence turnoff.    However, 
 \markcite{mlg90}Mathieu et al.\ (1990)
 found no evidence of a  secondary correlation peak
 in their high signal-to-noise spectra of S1040, indicating that the 
 secondary must be considerably fainter than the primary.      
 They also pointed out that the mass function was consistent with a secondary 
 mass as low as 0.18 \msun.
 To explain the circularization of
 the orbit, \markcite{vp95}Verbunt and Phinney (1995) suggested that the
 secondary of S1040 was a white dwarf, and that the white dwarf progenitor
 must have filled its Roche lobe.

 An important clue to the nature of the S1040 secondary was provided by a 
 1620 \AA\ image of M67 obtained in 1995 March with the 
 {\em Ultraviolet Imaging Telescope} (\cite{stech97}).    A total of 16 stars
 in M67 were detected at 1620 \AA, including the 11 hottest blue stragglers, 
 four white dwarf candidates, and S1040 (\cite{land97}).
 The detection of S1040 at 1620 \AA\ almost certainly implies that the 
 secondary is a hot white dwarf, since a more luminous type of ultraviolet
 source would be inconsistent with the composite red B--V color.       
 Since a white dwarf secondary makes a negligible contribution to the 
 integrated V magnitude, the peculiar location of S1040 in the M67 CMD is 
 likely the result of an earlier mass-transfer episode.
 To further elucidate the nature of S1040, we have now obtained
 observations of S1040 with the Goddard High-Resolution Spectrograph (GHRS) 
 and the Faint Object Spectrograph (FOS) aboard the {\em Hubble Space 
 Telescope}.

 The spectroscopic determination of the fundamental parameters of a 
 post-mass-transfer binary is often problematic, because both components 
 have deviated from single star evolution.   Thus the membership of S1040
 in the well-studied solar-metallicity open cluster M67 is particularly 
 fortunate, and we adopt
 in this {\em Letter} the cluster distance (820 pc), age (4 Gyr), 
 and reddening (E(B--V) = 0.025)
 given by \markcite{carr96}Carraro et al.\ (1996). 

\section{Observations}

S1040 was observed on 18 May 1996 with the G140L mode of the GHRS, 
covering the wavelength region from 1175 \AA\ to 1450 \AA, and with the 
G270H mode of the FOS, covering the region from 2220 \AA\ to 3300 \AA.
The observation date corresponds to phase 0.54 in the Mathieu et al.\ (1990) 
orbit.    
The exposure time of the GHRS observation was 6582s, and 
the use of the large ($1.74'' \times 1.74''$) aperture gave a spectral 
resolution of about 0.8 \AA.   
The exposure time of the FOS observation was 300s, and the use of the 
large ($3.7'' \times 1.3''$) aperture gave a 
spectral resolution of about 2.0 \AA.
Spectra from both
instruments were reduced using the software prepared by the GHRS instrument 
team (\cite{rob92}).   The GHRS spectrum was further corrected for a
$\sim 5$\% sensitivity degradation below 1200 \AA\ (\cite{sherb96}).

\section{Analysis}

The GHRS spectrum of S1040 (Figure 1) shows a broad \lya\ absorption 
that confirms the identification of the S1040 secondary as a hot white dwarf.
S1040 is among the faintest white dwarfs ever observed in the ultraviolet, and 
so despite the deep GHRS exposure, the S/N per resolution element is only 
about 18.  The strong emission in the core of \lya\ is due to the Earth's 
geocorona, and the strength 
and spectral profile of the emission feature near 1304 \AA\ 
is also consistent with being entirely due to diffuse  gecoronal \ion{O}{1} 
emission.   Interstellar lines of \ion{Si}{2} $\lambda$1260, 
\ion{C}{2} $\lambda$1335,
\ion{O}{1} $\lambda$1302, and \ion{Si}{2} $\lambda$1304 are clearly present,
while other possible absorption features at $\lambda$1180, $\lambda$1329, and 
$\lambda$1371 are of uncertain origin.  
The S/N and spectral resolution of the GHRS spectrum is insufficient to 
determine whether narrow photospheric lines exist in the white dwarf spectrum,
such as would be needed for an eventual determination of a 
double-lined spectroscopic orbit.

We fit the GHRS spectrum using the pure-hydrogen white dwarf model atmospheres 
of \markcite{berg95}Bergeron et al.\ (1995).   
A low-dispersion ultraviolet spectrum is not 
sufficient by itself to constrain both \teff\ and \logg\ in a hot white dwarf,
because a  good fit is possible at any value of \logg (\cite{lsb96}).
However, for an assumed value of \logg, the best-fit value of \teff\ and the
angular diameter derived from the flux scaling
can be used along with a (\teff-dependent) theoretical 
mass-radius relation to derive the white
dwarf distance.     Table 1 shows the computed distances to S1040 for a 
range of assumed \logg\ values, using the mass-radius relation of 
\markcite{wood95}Wood (1995) 
for carbon white dwarfs with thick hydrogen and helium layers.
Even for the lowest mass (0.2 \msun) carbon model 
available, the computed distance is less than the cluster distance
of 820 pc.     Qualitatively, the origin of this low 
mass determination is that a relatively cool \teff\ is required to fit the 
broad \lya\ absorption profile, so that a large radius (low mass) is needed to 
reproduce the absolute ultraviolet flux level.    

In fact, the use of a carbon composition to compute a mass-radius relation 
is not realistic for a such a low-mass white dwarf, 
because the ignition of helium  requires a core mass of at least 0.49 \msun 
near the tip of the red giant branch
(c.f.\ \cite{marsh95}).     At high ($\sim$ 50,000 K) temperatures,
a helium white dwarf is known to have a significantly larger radius than a 
carbon white dwarf of the same mass (\cite{venn95}), and some modifications in
the mass-radius relation might still be expected at the relevant \teff\ 
($\sim$16,000 K) for S1040.    Therefore, we have 
recomputed the derived distances in Table 1, using 
new evolutionary calculations of the helium-rich core of a giant star whose 
envelope has been stripped (see \cite{ap97} for more details).     The M67 
distance of 820 pc can then be matched using a model with \teff\ = 16,160 K, 
\logg\ = 6.7, and a mass of $\sim 0.224$ \msun.     
The evolutionary time for the white dwarf to reach this \teff\ is about 75 Myr 
after ejection of the envelope on the giant branch.    Note that, whereas the
evolutionary track of a typical 0.6 \msun\ white dwarf crosses into the hottest 
($\sim 10^5$ K) region of the HR diagram, the
maximum \teff\ along the 0.22 \msun\ evolutionary track is only about 17,500 K.

The FOS spectrum of S1040 is shown in Figure 2, along with the model white
dwarf spectrum derived from the fit of the GHRS spectrum.   The
white dwarf still dominates the spectrum at 2200 \AA, but provides only about 
1\% of the total flux at 3300 \AA.    Also shown in Figure 2 is a 
best-fit \markcite{kur93}Kurucz (1993) model with [Fe/H] = 0.0, \teff\ = 5150 K and 
\logg\ = 3.5, normalized to the reddening-corrected V magnitude of S1040.
The scaling required to fit the Kurucz model yields a radius of the yellow 
giant of 5.1 \rsun.    A radius about five times larger than this value would be
needed for S1040 to show an eclipse of the white dwarf, assuming the Mathieu 
et al.\ mass function, and masses of 1.5 \msun\ and 0.22 \msun for the primary
and secondary.

Figure 2 also shows that the \mgii $\lambda$2800 doublet is observed in 
emission, with an reddening-corrected integrated flux of 
$1.87 \times 10^{-14}$ erg cm$^{-2}$ s$^{-1}$.    This 
corresponds to a surface flux of $9.4 \times 10^{5}$ erg cm$^{-2}$ s$^{-1}$,
using the angular diameter derived above.
This \mgii\ surface flux is an order of magnitude larger than 
the typical values reported for seven M67 giants by \markcite{dup90}Dupree,
Hartman, \& Smith (1990), and near the upper envelope of the surface flux values
observed in field G giants.   The high chromospheric activity level 
indicated by this \mgii\ surface flux value, supports the classification of 
S1040 as an RS CVn binary, as originally suggested on the basis of its X-ray 
luminosity by \markcite{bell93}Belloni et al.\ (1993).

Table 2 summarizes the parameters of S1040, as derived in this work and
taken from the literature.

\section{Discussion}

The relatively long orbital period of S1040, and the low mass derived here for
the white dwarf secondary, place strong constraints on the possible
evolutionary history of S1040. 
An episode of mass transfer must have occured before the white dwarf 
progenitor developed a core mass which is larger 
than the mass of the current white dwarf.   
But if the mass transfer episode began while the donor still had a radiative 
envelope, then a common envelope would likely occur, leading to a shrinkage of 
the orbit, in contradiction to the observed 42.8d period.
In fact, population synthesis calculations (\cite{stef97}) indicate it should
not be rare for mass-transfer during the subgiant or lower giant branch of the
primary (early Case B evolution) to lead to 
systems similar to S1040.  
For such systems, a relation
between the final orbital period  and the mass of the helium 
white dwarf can be derived by combining the core mass -- radius 
relation for a red 
giant with the requirement that the red giant always fill its 
Roche lobe  during the mass-transfer phase (e.g.\ \cite{rapp95}).      
The Rappaport et al.\ 
relation predicts a white dwarf mass of $\sim 0.27$ \msun for the S1040 orbital
period, while a similar formula of Eggleton (personal communication) 
predicts a white dwarf 
mass of 0.25 \msun.     These predicted masses are sufficiently close to the
white dwarf mass of 0.224 \msun\ derived here, to lend confidence in our ability
to outline the evolutionary history of S1040.

The basic scenario is that two stars in a relatively close orbit 
begin an epoch of
mass transfer when the more massive star fills its Roche lobe.
If the initial orbital period is on the order of days, the
donor star will have a helium core mass near or slightly
larger than $0.1 M_\odot$ when mass transfer begins. 
Until sometime after the masses
equalize, the mass transfer rate will be fairly high, governed 
by the the donor's attempts to re-establish thermal equilibrium 
as it expands, while its Roche lobe either shrinks
or remains close to a fixed volume. After mass equalization,
continued mass transfer tends to expand the Roche lobe, even as
the donor itself expands, leading to an epoch of stable mass transfer.
The mass transfer rate during
this latter epoch is governed by the donor's nuclear evolution time scale,
which can be longer than it would have been had the donor mass remained 
constant. Mass transfer ends when the donor's envelope is
depleted, leaving the white dwarf remnant we observe.

For concreteness, we have constructed a binary evolution model which yields a 
system with the approximate parameters derived here for S1040.  
The final white dwarf mass (0.25 \msun) and age (5.0 Gyr) in the model
are slightly    
larger than values reported here for S1040, but probably within the
observational errors. 
The model ingredients are described by \markcite{stef96}Di Stefano \&
Nelson (1996) and \markcite{stef97}Di Stefano (1997), and we note here only that
the mass retention factor,
$\beta$, is taken to be proportional to the ratio of the donor's thermal time 
scale to  that of the accretor. 
The initial model masses are 1.24 and 0.82 \msun, and the initial orbital
period is 1.86d.      For such a system, Roche lobe overflow occurs after about
4.1 Gyr, when the more massive star has developed a core mass of 0.159
\msun.     After 95 Myr, the 
component masses have equalized at about 0.98 \msun (with about 0.1 \msun\ lost
from the system), although the orbital period
(1.97d), and the donor core 
mass (0.163 \msun), have increased only slightly.
Stable mass transfer continues for another 765 Myr until the envelope 
of the donor is depleted, leaving a system with an orbital period of 42.1 d,
consisting of a 0.25 \msun\ helium white dwarf and a 1.48 \msun\  blue
straggler.    In fact, the star is observed as a blue straggler   
(i.e.\ the accretor mass exceeds the cluster turnoff mass) 
during the final $\sim 400$ Myr of mass transfer.

As noted earlier, comparison with the 0.22 \msun\
evolutionary track indicates that mass transfer in S1040 ended 
about 75 Myr ago.   Thus, within the past 
75 Myr the blue straggler in S1040 must have evolved redward to its current 
location, 0.2 mag blueward of the giant branch in the M67 CMD.    Qualitatively,
an evolved blue straggler is expected to lie to the blue of the cluster giant
branch, but a more detailed calculation than presented in this {\em
Letter} will be needed to follow the temperature and luminosity evolution of
the accretor. 

Are there other systems similar to S1040?  
Case B mass transfer is the most likely explanation for the 
origin of the current M67 blue straggler F190, which has a 4.2d period, 
although it may not be possible to explain the other M67 
blue stragglers as the end products of mass transfer (\cite{leonard96}).   
Among the field stars,
many of the parameters of AY Cet (= HR 373, K0 IV + wd, P = 56.8d) are 
similar to those of S1040, including the mass function and the estimated 
white dwarf temperature (\cite{sim85}).    
Simon et al.\ point out that the
white dwarf mass in AY Cet could be as low as 0.25 \msun, provided that its 
distance is near the the upper limit of that derived from the 
(large) uncertainty in its parallax.    The hot component in the eclipsing 
binary HD 185510 (K0 III/IV, P = 20.7d) is known to have a low (0.3 \msun) mass, 
but whether its atmospheric parameters are consistent with those of a 
degenerate helium white dwarf is still uncertain (\cite{jeff97}).

\acknowledgments

Support for this work was provided by NASA through grant number 
GO-06680.01.94A from the
Space Telescope Science Institute, which is operated by the Association of
Universities for Research in Astronomy, Inc., under NASA contract NAS5-26555.
JMA acknowledges financial support the DGICYT project PB94-0111 (Spain)
and a CIRIT grant (Catalonia).    We thank Doyle Hall, Allen Sweigart, and 
Ben Dorman for useful discussions, and Matt Wood for his white dwarf cooling 
tracks.

\clearpage

\begin{figure}
\plotone{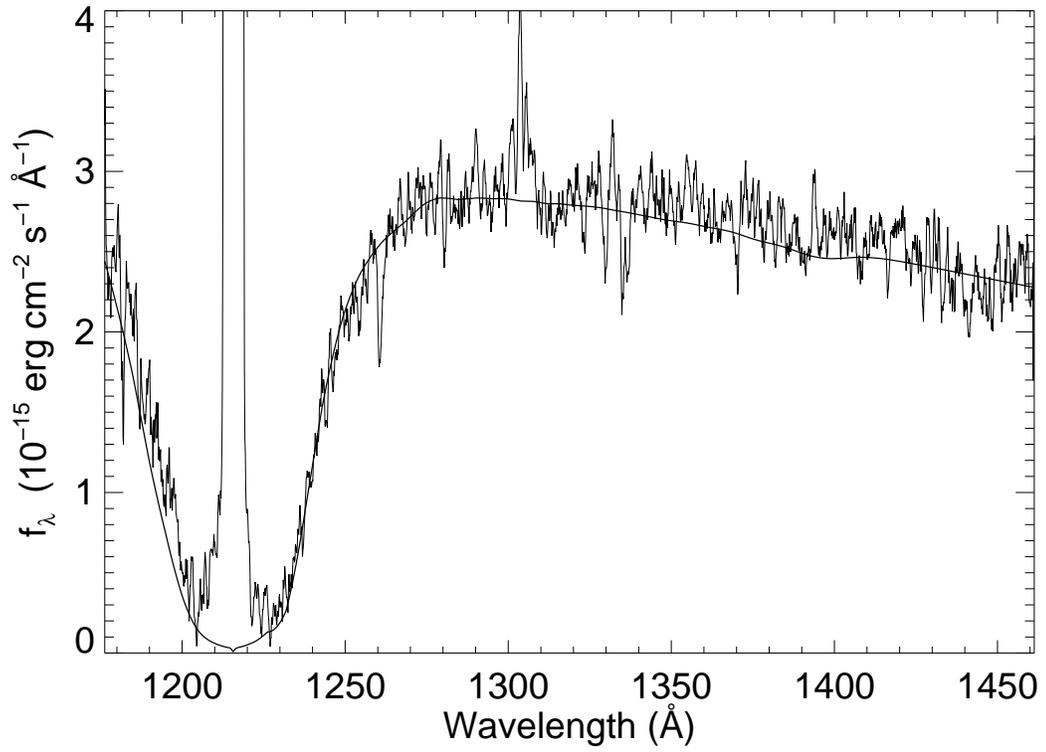}
\caption{GHRS spectrum of S1040.  The spectrum has been corrected for a 
reddening of E(B--V) = 0.02.    The thick solid line shows a best-fit
white dwarf model with \teff\ = 16,900 K and \logg = 7.0.}  

\end{figure}

\clearpage

\begin{figure}
\plotone{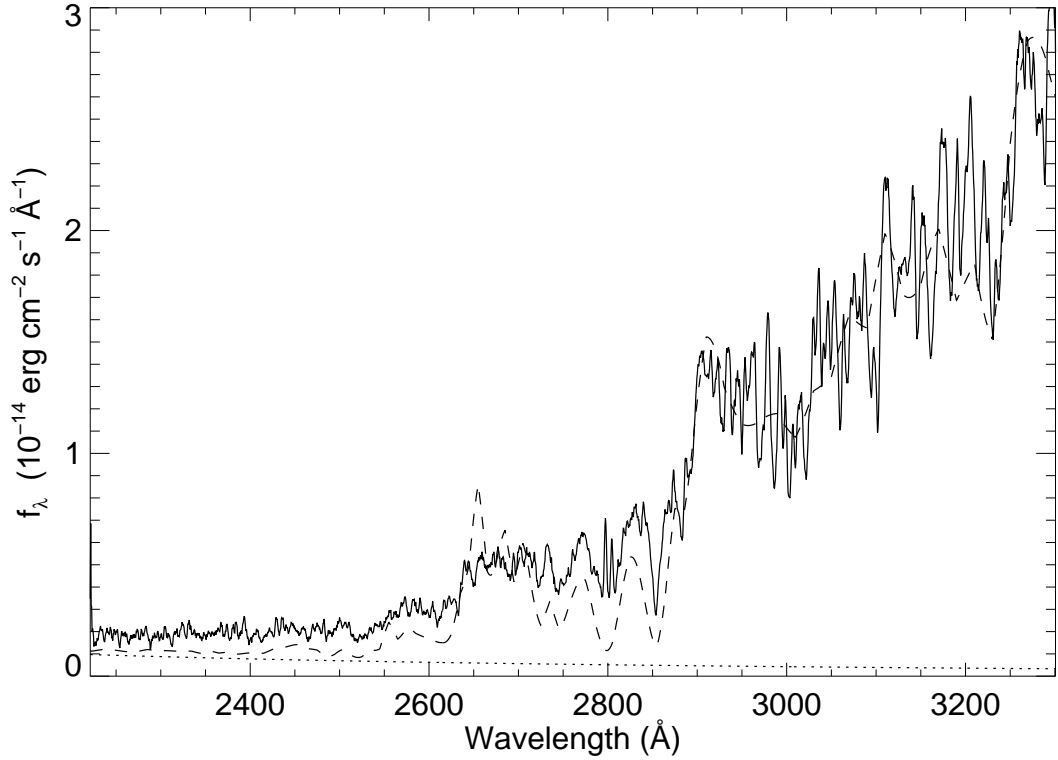}
\caption{FOS spectrum of S1040.  The spectrum has been dereddened by
E(B--V) = 0.02.   The dotted line shows the predicted 
contribution of the white dwarf, using the parameters derived from 
the fit of the GHRS spectrum.   The dashed line shows the sum of the 
white dwarf spectrum and a Kurucz model with \teff\ = 5150 K, \logg = 3.5,
normalized to the reddening-corrected V magnitude of S1040.}

\end{figure}
\clearpage

\tightenlines

\begin{deluxetable}{llllll}  
\tablecaption{White-Dwarf Model Fits}
\tablehead{ \colhead{\logg}  & \colhead{\teff} 
   & \colhead{R$^2$/D$^2$} & \colhead{M/\msun} & \colhead{{\em d} (pc)} &
Compos }
\startdata
 6.7 & 16,135 & $9.06 \times 10^{-25}$ & 0.223  & 825 & He \\
     &        &                        & 0.200  & 791 & C  \\
 7.0 & 16,890 & $7.19 \times 10^{-25}$ & 0.262  & 718 & He \\
     &        &                        & 0.254  & 714 & C  \\ 
 7.5 & 18,740 & $4.35 \times 10^{-25}$ & 0.39   & 645 & C \\
 8.0 & 20,660 & $2.75 \times 10^{-25}$ & 0.62   & 567  & C \\
\enddata                                      

\end{deluxetable}

\begin{deluxetable}{lll}  
\tablecaption{Parameters of S1040}
\tablehead{\colhead{Parameter}  & \colhead{Value} & 
 \colhead{Ref\tablenotemark{a}}  }
\startdata
 V & 11.51 & 1  \\
 B--V & 0.86 & 1  \\
 Sp.T. & G4 III & 2 \\
 Period (d) & 42.8 & 3 \\
 K (km s$^{-1}$) & 8.45 & 3 \\
 f(m) & 0.00268 & 3 \\
 L$_{X}$ (erg s$^{-1}$) & $3 \times 10^{30}$  & 4 \\
 \teff (primary) & 5150 K  & 5 \\
 R (primary)  & 5.1 \rsun\ & 5   \\
 M (WD) & 0.22 \msun\ & 5 \\
 \teff (WD) &  16,160 K & 5 \\
\enddata

\tablenotetext{a}{REFERENCES: (1) \cite{js84}, 
(2) \cite{as95}, (3) \markcite{mlg90}Mathieu et al.\ 1990, 
(4) \markcite{bell93}Belloni et al.\ 1993, (5) this work}

\end{deluxetable}

\end{document}